\documentclass{appolb}
\usepackage{epsfig}

\begin{document}

\title{Recent Experimental Results on Nuclear Cluster Physics}

\author{C. Beck$^a$
\address{
$^a$D\'epartement de Recherches Subatomiques, Institut Pluridisciplinaire 
Hubert Curien, IN$_{2}$P$_{3}$-CNRS and Universit\'e de Strasbourg - \\
23, rue du Loess BP 28, F-67037 Strasbourg Cedex 2, France\\
E-mail: christian.beck@iphc.cnrs.fr\\}
}

\maketitle

\begin{abstract}
Knowledge on nuclear cluster physics has increased considerably since the 
pioneering discovery of $^{12}$C+$^{12}$C resonances half a century ago and 
nuclear clustering remains one of the most fruitful domains of nuclear physics, 
facing some of the greatest challenges and opportunities in the years ahead. 
The occurrence of ``exotic" shapes and/or Bose-Einstein $\alpha$ condensates 
in light $N$=$Z$ $\alpha$-conjugate nuclei is investigated. Evolution of 
clustering from stability to the drip-lines examined with clustering aspects 
persisting in light neutron-rich nuclei is consistent with the extension of 
the ''Ikeda-diagram" to non $\alpha$-conjugate nuclei.

\end{abstract}

\newpage

\section{Introduction}
\label{sec:1}

Among the greatest challenges in nuclear science was the understanding of the clustered 
structure of nuclei from both the experimental and theoretical perspectives
\cite{Greiner95,Cluster1,Cluster2,Cluster3,Delion,Freer14,Funaki15,Beck17}. Progress in  
physics of nuclear molecules and nuclear clustering were facing some of the greatest 
challenges and opportunities in the years ahead. Besides the well known series of Cluster 
conferences~\cite{Nara,Stratford,Debrecen}, a series of workshops on the State-Of-The-Art 
in Nuclear Cluster Physics was started~cite{Strasbourg,Brussels,Yokohama}. The first one 
was held in Strasbourg in 2008~\cite{Strasbourg}, the second one in Brussels in 
2010~\cite{Brussels} and the last one in Yokohama in 2014~\cite{Yokohama}. Fig.~1 (taken 
from the cover of Ref.~\cite{Yokohama}) summarizes the different types of clustering that 
were discussed during the last two or three decades~\cite{Beck17,Beck15}. Most of these 
exotic structures were investigated in an experimental context by using either some new 
approaches or developments of older methods~\cite{Papka12}.\\

Starting in the 1960s the search for resonant structures in the 
excitation functions for various combinations of light $\alpha$-conjugate ($N$=$Z$) nuclei 
in the energy regime from the Coulomb barrier up to regions with high excitation energies 
($E_{x}$=20$-$50~MeV) remains a subject of contemporary debate~\cite{Greiner95,Beck17}. These 
resonances were interpreted in terms of nuclear molecules~\cite{Greiner95}. 
The question of how quasimolecular resonances may reflect continuous transitions
from scattering states in the ion-ion potential to true cluster states in the 
compound systems was still unresolved in the 1990s \cite{Greiner95}. 
In many cases, these 
resonant structures were associated with strongly-deformed shapes and with $\alpha$-clustering 
phenomena~\cite{Freer14,Beck17,Freer07,Horiuchi10}, predicted from the Nilsson-Strutinsky approach, the 
cranked $\alpha$-cluster model~\cite{Freer14,Freer07}, or other mean-field 
calculations~\cite{Horiuchi10,Gupta10}. \\

In light $\alpha$-like nuclei clustering is 
observed as a general phenomenon not only at high excitation energies close to the 
$\alpha$-decay 
thresholds~\cite{Freer14,Beck17,Freer07,Oertzen06} but may also play a key role at low 
excitation energies and for the ground states. This exotic behavior was perfectly illustrated by the 
famous ''Ikeda-diagram" for $N$=$Z$ nuclei in 1968 \cite{Ikeda}, which was modified 
and recently extended by von Oertzen \cite{Oertzen01,Milin14} for neutron-rich nuclei, as 
shown in the right panel of Fig.~2. Despite the early inception of cluster studies, it is 
only recently that radioactive ion beams experiments, with great help from advanced 
theoretical works, enabled new generation of studies, in which data with variable excess 
neutron numbers or decay thresholds are compared to predictions with least or no assumptions 
of cluster cores. For instance, new experimental approaches for the tetraneutron system at 
RIKEN \cite{Kisamori16} challenge the most advanced theoretical models
\cite{Rimas16,Bertulani16}. On the other hand, predicted but elusive phenomena, such as 
molecular orbitals or linear chain structures, are now gradually coming to light from
experimental data~\cite{Freer14,Beck17}.  \\

Clustering is a general feature \cite{Freer14,Beck17,Milin14} not only observed in light 
neutron-rich nuclei \cite{Kanada10}, but also in halo nuclei  such as $^{11}$Li \cite{Ikeda10} 
or $^{14}$Be \cite{Nakamura12}, for instance. The problem of cluster formation has also been 
treated extensively for very heavy systems by R.G. Gupta \cite{Gupta10}, by D. Poenaru, 
V. Zagrebaev and W. Greiner \cite{Poenaru10,Zagrebaev10} and by C. Simenel \cite{Simenel14} 
suggesting that giant molecules and collinear ternary fission may co-exist \cite{Kamanin14}. 
Finally, signatures of $\alpha$-clustering have also been predicted and/or discovered in light 
nuclei surviving from intermediate-energy \cite{Borderie16} to ultrarelativistic-energy
\cite{Zarubin14} nuclear collisions. The topic of nuclear clusters benefits of intense 
theoretical activity~\cite{Funaki15} where new experimental information has come to light 
very recently \cite{Freer14,Beck17}. Several status reports were given in conferences and 
their written contributions can be found in their respective 
proceedings~\cite{Beck15,Beck13a,Beck16}.

\begin{figure}[th]
\centerline{\psfig{figure=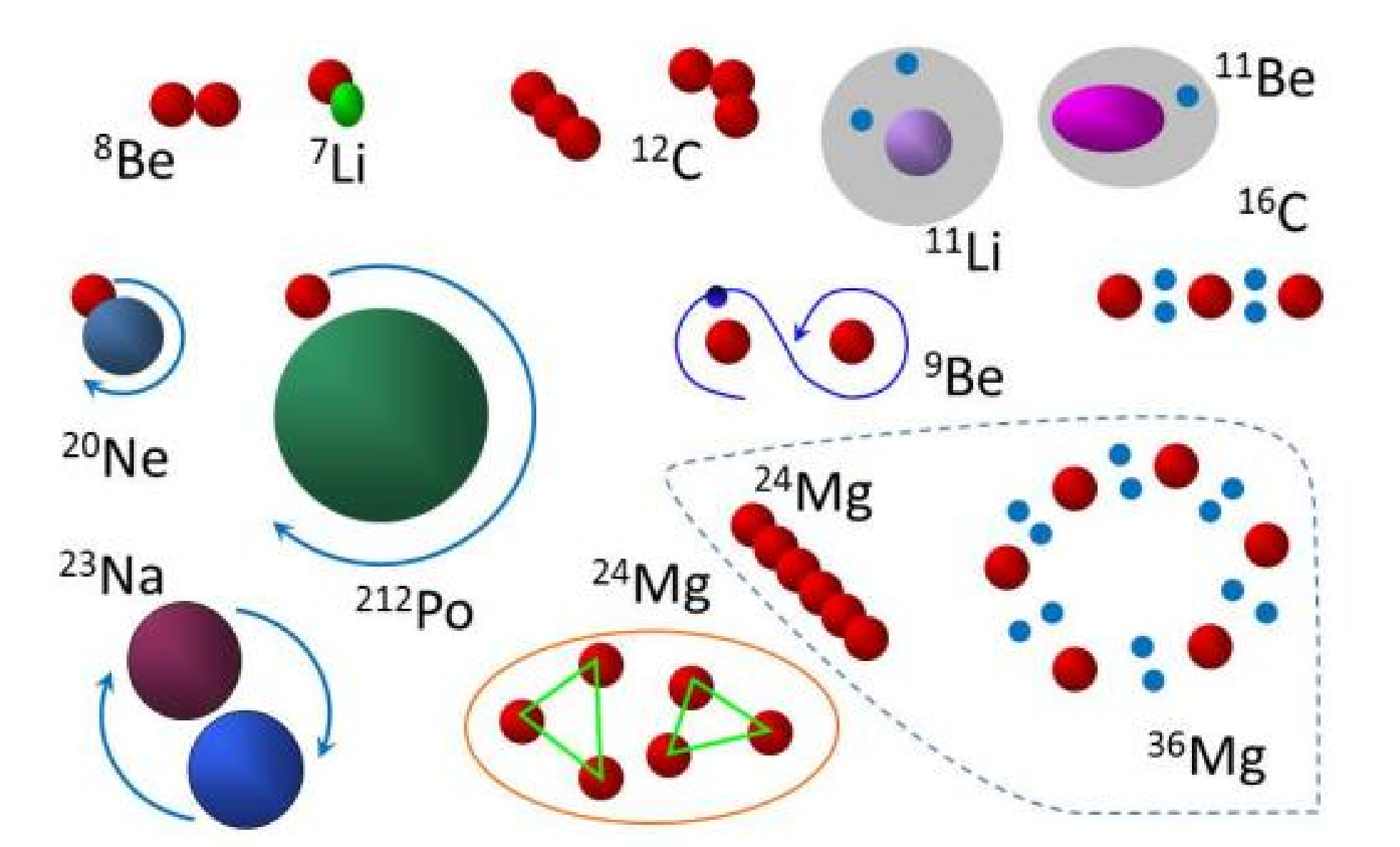,width=13.2cm,height=8.4cm}}
\caption{\label{fig1} Different types of nuclear clusters
discussed in recent years \cite{Beck17,Beck15,Beck13a,Beck16}.}
\vspace*{-10pt}
\end{figure}

\newpage

\section{Renewed interest in the spectroscopy of light $\alpha$-like nuclei: $\alpha$
condensates ?}
\label{sec:2}

The renewed interest in the $^{12}$C nucleus~\cite{Freer14a} was mainly focused to a better 
understanding of the nature of the so called "Hoyle" state \cite{Hoyle54} that can be described 
in terms of a bosonic condensate, a cluster state and/or a $\alpha$-particle gas 
\cite{Tohsaki01,Oertzen10a,Yamada}. Much experimental progress has been achieved recently as 
far as the spectroscopy of  $^{12}$C near and above the $\alpha$-decay threshold is
concerned~\cite{Freer14a,Zimmerman13,Marin14}. More particularly, the 2$^{+}_{2}$ "Hoyle" 
rotational excitation in  $^{12}$C has been observed \cite{Zimmerman13}. The 
$^{12}$C($\alpha$,$\alpha$)$^{12}$C$^*$ reaction~\cite{Marin14} populates a new state compatible 
with an equilateral triangle configuration of three $\alpha$ particles. Still, the structure 
of the "Hoyle" state remains controversial as experimental results of its direct decay into 
three $\alpha$ particles are found to be in disagreement ~\cite{Freer14a,Itoh14}.\\

In the study of Bose-Einstein Condensation (BEC), the $\alpha$-particle states in light 
$N$=$Z$ nuclei \cite{Tohsaki01,Oertzen10a,Yamada}, are of great importance. At present, the 
search for an experimental signature of BEC in $^{16}$O is of highest priority. 
A state with the structure of the ''Hoyle" state \cite{Hoyle54} in $^{12}$C coupled to an 
$\alpha$ particle is predicted in  $^{16}$O at about 15.1 MeV (the 0$^{+}_{6}$ state), the
energy of which is $\approx$ 700 keV above the 4$\alpha$-particle breakup threshold 
\cite{Funaki08}. However, any state in $^{16}$O equivalent to the ''Hoyle" state \cite{Hoyle54} 
in $^{12}$C is most certainly going to decay exclusively by particle emission with very small 
$\gamma$-decay branches, thus, very efficient particle-$\gamma$ coincidence techniques will 
have to be used in the near future to search for them. \\

BEC states are expected to decay by 
$\alpha$ emission to the ''Hoyle" state and could be found among the resonances in 
$\alpha$-particle inelastic scattering on $^{12}$C decaying to that state. In 1967, Chevallier 
et al.~\cite{Chevallier67} could excite these states in an $\alpha$-particle transfer channel 
leading to the $^{8}$Be--$^{8}$Be final state and proposed that a structure corresponding to 
a rigidly rotating linear arrangement of four alpha particles may exist in $^{16}$O.
A more sophisticated experimental setup was used recently \cite{Curtis13}: although the excitation function is
generally in good agreement with the previous results \cite{Chevallier67} a phase shift analysis 
of the angular distributions does not provide evidence to support the reported hypothesis of a 
4$\alpha$-chain state configuration. Tetrahedral symmetries are predicted to occur in $^{16}$O 
\cite{Iachello14}. Experimental investigations are still underway to understand the nuclear 
structure of high-spin states of $^{16}$O, $^{20}$Ne and heavier nuclei
\cite{Kokalova13b,Papka14}. The search for exotic chain-like structures remains an exciting 
prospect but, up to now, tentative evidence of chain states have been unsubstantiated
and the view is that such structure have not yet been definitively observed.

\newpage

\section{Alpha clustering, deformations and $\alpha$ condensates in heavier nuclei}
\label{sec:3}

The relashionship between $\alpha$-clustering, nuclear molecules and superdeformation (SD) 
\cite{Beck17,Horiuchi10,Beck04a,Ray16} is of particular interest, since nuclear shapes with 
major-to-minor axis ratios of 2:1 are typical ellipsoidal elongations for light nuclei 
(corresponding to a quadrupole deformation parameter $\beta_2$ $\approx$ 0.6). Furthermore, 
the structure of possible octupole-unstable 3:1 nuclear shapes (hyperdeformation (HD) 
with $\beta_2$ $\approx$ 1.0) has also been discussed for actinide nuclei in terms of 
clustering phenomena \cite{Beck17}. Typical examples for the link between quasimolecular 
bands and extremely deformed (SD/HD) shapes can be found in the literature for $A = 20-60$ 
$\alpha$-conjugate $N$=$Z$ nuclei 
\cite{Beck17,Horiuchi10,Beck13a,Beck04a,Jenkins13,Jenkins15,Chandana10,Svensson00,Beck09,Ideguchi01,Rousseau02,Salsac08,DiNitto16,Nouicer99,Sciani09,Beck01,Bhattacharya02,Oertzen08}.
Highly deformed shapes and SD rotational bands have been discovered in several 
light $\alpha$-like nuclei, such as $^{36}$Ar and $^{40}$Ca 
by using $\gamma$-ray spectroscopy techniques \cite{Svensson00,Ideguchi01}. 
In particular, the extremely deformed rotational SD band in $^{36}$Ar 
\cite{Svensson00} is comparable in shape to the quasimolecular bands 
observed in both $^{12}$C+$^{24}$Mg and $^{16}$O+$^{20}$Ne reactions 
\cite{Beck17,Beck09,Sciani09}.\\
 
Ternary clusterizations are also predicted theoretically, but were not found 
experimentally in $^{36}$Ar so far \cite{Beck17,Beck09}. On the other hand, ternary 
fission of $^{56}$Ni, related to its HD shapes, was identified 
from out-of-plane angular correlations measured in the $^{32}$S+$^{24}$Mg 
reaction with the Binary Reaction Spectrometer at the {\sc Vivitron} Tandem 
facility \cite{Oertzen08}. This possibility \cite{Beck17,Oertzen08} is not limited 
to light $N$=$Z$ compound nuclei, true ternary fission 
\cite{Zagrebaev10,Kamanin14,Pyatkov10} can also occur for very heavy 
\cite{Kamanin14,Pyatkov10} and superheavy \cite{Zagrebaev10b} nuclei.
The next natural question to be addressed is whether
dilute-gas-like BEC structures~\cite{Tohsaki01,Oertzen10a,Yamada}
also exist in medium-mass $\alpha$-conjugate nuclei as predicted by several
theoretical investigations~\cite{Yamada04,Khan,Girod}. Several recent undergoing
experiments indicate that it might be the case at least for $^{24}$Mg
\cite{Francesca14}, and much work is in progress
in this field \cite{Beck17,Oertzen10a}.

\newpage

\section{Clustering in light neutron-rich nuclei}
\label{sec:3}

\begin{figure}[th]
\centerline{\psfig{figure=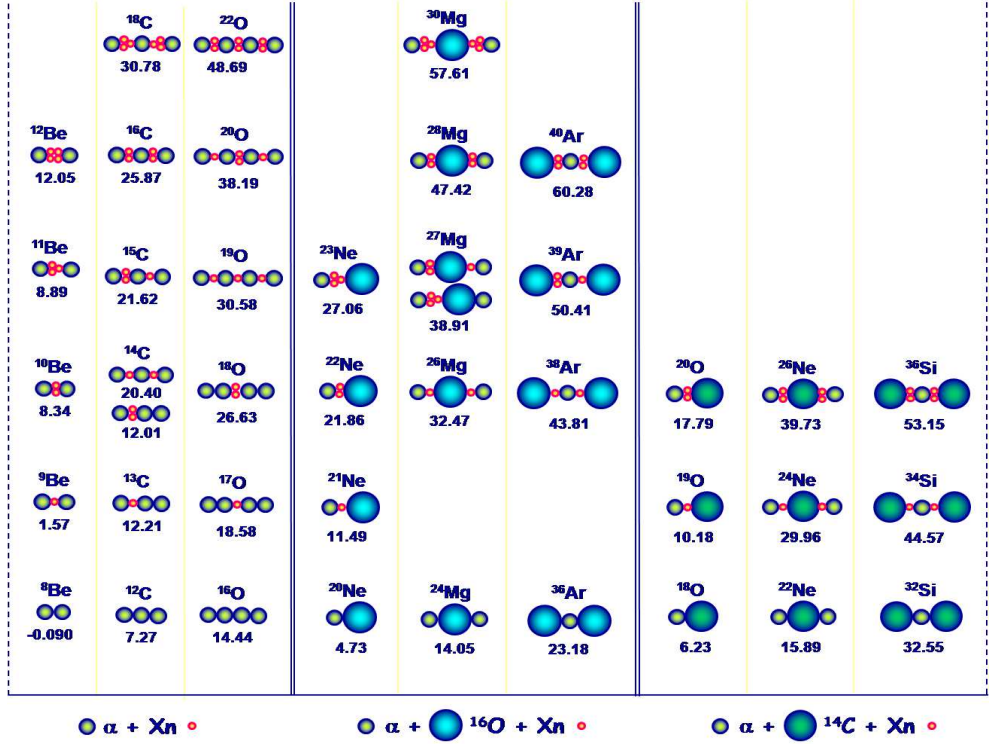,width=12.9cm,height=9.9cm}}
\vspace*{8pt}
\caption{\label{fig2} 
Schematic illustration of the structures of molecular
shape isomers in light neutron-rich isotopes of nuclei consisting
of $\alpha$-particles, $^{16}$O- and $^{14}$C-clusters plus some
covalently bound neutrons (Xn means X neutrons). The so called "Extended 
Ikeda-Diagram" \cite{Oertzen01} with $\alpha$-particles (left panel) and 
$^{16}$O-cores (middle panel) can be generalized to $^{14}$C-cluster cores 
(right panel). The lowest line of each configuration corresponds to parts
of the original "Ikeda-Diagram" \cite{Ikeda}.}
\end{figure}

Clustering is a general phenomenon \cite{Beck17} observed also in nuclei with extra neutrons 
as illustrated in Fig.~2 by the "Ikeda-Diagram" \cite{Ikeda} modified and 
extended by von Oertzen \cite{Oertzen01}. With additional neutrons, specific 
molecular structures appear with binding effects based on covalent molecular 
neutron orbitals. In these diagrams $\alpha$-clusters and 
$^{16}$O-clusters are the main ingredients. Actually, the $^{14}$C nucleus may 
play a similar role in clusterisation as the $^{16}$O
nucleus does. Both of them have similar properties as a cluster: i) closed 
neutron p-shells, ii) first excited states well above 
E$^{*}$ = 6 MeV, and iii) high binding energies for $\alpha$-particles.
A general picture of clustering and molecular configurations in light nuclei 
can be drawn from investigation of the oxygen isotopes with
A $\geq$ 17. Recent results on the even-even 
oxygen isotopes i.e. $^{18}$O \cite{Oertzen10b,Avila14} and 
$^{20}$O \cite{Milin14} 
as well as very striking cluster states found in odd-even oxygen 
isotopes such as $^{17}$O \cite{Milin09} and $^{19}$O \cite{Milin14}
have been obtained. 
Therefore, the "Ikeda-Diagram" 
\cite{Ikeda} and the "Extended Ikeda-Diagram" consisting of $^{16}$O cluster
cores with covalently bound neutrons \cite{Oertzen01} is
being further revised to 
include also the $^{14}$C cluster cores as illustrated in
Fig. 2 (right panel). 

\newpage

\section{Concluding remarks}

Marked progress is being accomplished in many traditional or novel subjects of nuclear
cluster physics. High-precision spectroscopy techniques enable us to uncover 
important parts of the complete spectroscopy of the "Hoyle" state in $^{12}$C. 
Thus, the origin of carbon for life is likely to be understood in the very near 
future with answer to the question of the "Hoyle" state structure.
The connection of $\alpha$-clustering and quasimolecular resonances with 
$\alpha$ condensates in very light nuclei and with extreme deformations 
(SD, HD, ...) in heavier nuclei as investigated by more and more sophisticated 
experimental devices is discussed in this introductory talk. 
Some neutron-rich nuclei displaying very well defined quasimolecular bands in 
agreement with theoretical predictions justify the "Extended Ikeda-Diagram" to 
be generalized by the inclusion of the $^{14}$C cluster as a core, similarly to the 
$^{16}$O one. The developments in these selected subjects of nuclear clusters 
physics show the importance of clustering among the basic modes of motion of 
nuclear many-body systems.

\newpage

\section{Dedication and acknowledments}

This article is dedicated to the memory of my friends Paulo Gomes, Alex Szanto 
de Toledo and Valery Zagrebaev. I would like to thank the CLUSTERS '16 
organizers for giving me the opportunity to introduce this conference. 
Christian Caron (Springer) is acknowledged for initiating the series of three 
volumes of \emph{Lecture Notes in Physics} entitled "Clusters in Nuclei". 

\newpage

\bibliographystyle{unsrt}
\bibliography{Napoli}

\end{document}